# Superconductivity in nickel based 112 systems

Qiangqian Gu*, Hai-Hu Wen‡

National Laboratory of Solid State Microstructures and Department of Physics,

Collaborative Innovation Center of Advanced Microstructures, Nanjing University,

Nanjing 210093, China

Qiangqiang Gu now is in Cornell University as a post-doctor. Corresponding authors qg55@cornell.edu, hhwen@nju.edu.cn

**Graphical abstract**

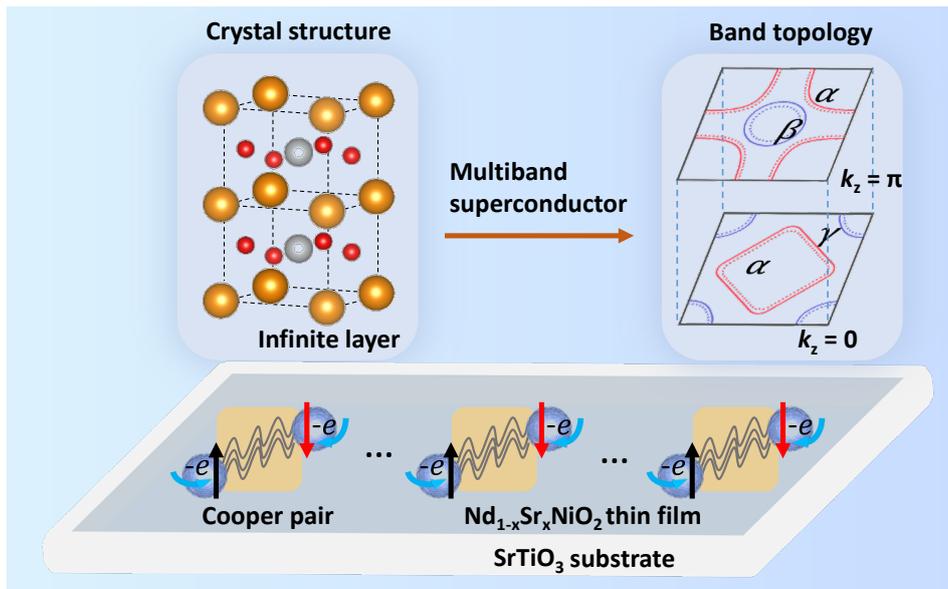

**Public summary**
1. The newly found nickel based 112 infinite thin films $R_{1-x}A_x NiO_2$ ($R$ = La, Nd, Pr and $A$ = Sr, Ca) can host superconductivity up to 15 K
2. $R NiO_2$ is a multiband system due to the hybridization between itinerant $R$ 5$d$ and Ni 3$d$ orbitals
3. The remaining antiferromagnetic fluctuations instead of long-range magnetic order can be detected in nickelate system
4. $R_{1-x}A_x NiO_2$ has unconventional pairing sate with a robust $d$-wave superconducting gap, and a full gap without unified understanding
5. The nickelate system provides a new platform for researching and exploring unconventional superconductivity

**ABSTRACT**

Superconductivity has been discovered recently in nickel based 112 infinite thin films $R_{1-x}A_x$NiO$_2$ (*R* = La, Nd, Pr and *A* = Sr, Ca). They are isostructural to the infinite-layer cuprate (Ca,Sr)CuO$_2$ and are supposed to have a formal Ni 3$d^9$ valence, thus providing a new platform to study the unconventional pairing mechanism of high-temperature superconductors. This important discovery immediately triggers a huge amount of innovative scientific curiosity in the field. In this paper, we try to give an overview of the recent research progress on the newly found superconducting nickelate systems, both from experimental and theoretical aspects. We will focus mainly on the electronic structures, magnetic excitations, phase diagrams, superconducting gaps and finally make some open discussions for possible pairing symmetries in Ni based 112 systems.

**INTRODUCTION**

In 1986, Bednorz and Müller in IBM Zürich Research Laboratory discovered the high temperature superconductivity in Ba doped insulating cuprate system La$_{2-x}$Ba$_x$CuO$_4$,[1] and they were awarded the Nobel Prize in Physics for this important discovery. Soon after, the critical temperature (*T*$_c$) in YBa$_2$Cu$_3$O$_{7-\delta}$ system was found to be as high as 93 K[2,3] which broke through the Macmillan limit[4] and the boiling temperature of liquid nitrogen, starting a new era of intensive research in high-temperature superconductivity (HTS). People have discovered more and more cuprate systems[5-8] and continuously renewed the records of the highest superconducting transition temperatures at ambient pressure.[9,10] However, the pairing mechanism in

HTS remains controversial and elusive.[11-13] It seems that the superconducting states can coexist or compete with different kinds of emergent intertwined orders leading to a complicated phase diagram.[14-18] Since 2008, the discovery of Fe-based superconductors (FeSC)[19,20] gives us some new hints for researching the pairing mechanism of HTS and exploring superconductors with higher $T_c$. Inspiringly, Chinese scientists have made great contributions to the fast developing field of superconductivity during the last decades.[21-23] For now, cuprate and FeSC are recognized as the only two systems of HTS under ambient pressure and it seems to be a long-standing pursuit to realize room temperature superconductors.[24-27] The most essential consensus has been achieved is that HTS should have layered structure and 3$d$ orbital electrons of transition metals with appropriate electron correlation, evolving from antiferromagnetic (AF) spin fluctuation or superexchange effect.

Many experimental and theoretical efforts have been devoted to searching HTS in the analogues without copper. For example, bulk superconducting $Sr_2RuO_4$ has very low $T_c$[28] and electron-doped $Sr_2IrO_4$ exhibits plausible spectroscopic signatures of a superconducting gap.[29] As shown in Figure 1A, the element of Ni locates just between Fe and Cu, and we may wonder whether HTS can be realized in the Ni based materials. In this regard, $LaNiO_3/LaAlO_3$ superlattice has been proposed as a potential candidate for high temperature superconductivity,[30-32] here Ni has a 3$d^7$ orbital configurations and the $d_{x^2-y^2}$ orbital is lowered in energy below the $d_{z^2}$ orbital. This will lead to a half-filled $d_{x^2-y^2}$ orbital, mimicking the electron occupation of undoped cuprates. However, the most stable nickelates in nature have a formal valence of $Ni^{2+}$ with a $d^8$ electronic configuration, such as in NiO and $La_2NiO_4$. In spite of the same structure as $La_2CuO_4$, people failed to find any trace of superconductivity in the hole doped system $La_{2-x}Sr_xNiO_4$.[33,34] Additionally, the multilayer nickelate Ruddlesden-Popper phase[35,36] $R_{n+1}Ni_nO_{3n+1}$ ($R$ = La, Nd, Pr) possess high valence state of Ni. The apical oxygen can be removed using a soft-chemistry topotactic reduction method, resulting in the phase of $R_{n+1}Ni_nO_{2(n+1)}$ with a relatively lower valence state.[37-39] Their Ni $3d_{x^2-y^2}$ band is expected to be close to half filling as the number of layers n increases.[40] However, people found some other exotic physics involving charge and spin orders instead of superconductivity in this family of materials.[40-43] In particular, when n approaches to infinity, it can be reduced to infinite-layered $RNiO_2$ with $3d^9$ configuration of $Ni^{1+}$ valence.[44,45] With the dedicated efforts in decades, in 2019, superconductivity was finally discovered in $Nd_{1-x}Sr_xNiO_2$ (x = 0.2) thin films deposited on $SrTiO_3$ substrates.[46] Figure 1B shows that the apical oxygen of $Nd_{1-x}Sr_xNiO_3$ can be removed by using $CaH_2$ reduction at high temperature, forming a superconducting phase $Nd_{1-x}Sr_xNiO_2$. Bulk $NdNiO_3$ is a nearly standard perovskite structure with room-temperature lattice parameters 3.81 Å and Sr doping gives negligible influence on the lattice parameter.[47] The reduction will lead to an expansion of the in-plane lattice constants as about 3.92 Å and a reduced c axis of about 3.31 Å, which is the distance between adjacent $NiO_2$ planes.[38] $NdNiO_2$ is composed by the stacking of alternating $NiO_2$ plane (superconducting layer) and Nd plane (charge reservoir), forming this so-called infinite layer structure. In the resistivity measurements, undoped $NdNiO_3$ shows the characteristic first order phase transition from a high-temperature paramagnetic metal to a low-temperature antiferromagnetic insulator.[48,49] Sr doping makes it change into a good metal in the whole temperature region. After the reduction, NdNiO$_2$ displays metallic behavior at high temperatures, with a resistive upturn below about 70 K. This unexpected feature will be discussed in the following section. Upon 20% Sr doping, Nd$_{0.8}$Sr$_{0.2}$NiO$_2$ exhibits a superconducting transition with an onset at 14.9 K and zero resistance at 9.1 K. Furthermore, perfect diamagnetism was observed recently on Nd$_{0.8}$Sr$_{0.2}$NiO$_2$ thin films with different thickness ranging from 6 to 17 nm.[50] Figure 1C shows the DC magnetization measured under zero field cooling (ZFC) and field cooling (FC) modes on the thin film with typical thickness of 7 nm. It is found that $T_c$ of diamagnetism is slightly lower than the one of zero resistance, suggesting the presence of inhomogeneity of the superconducting phase in the thin films. The phase coherence can only occur in the entire films when it is well below the zero-resistance temperature, and the Meissner effect can thus happen. Combined with the fact of considerable critical current density up to 170 kA cm$^{-2}$,[46] we may conclude that the Nd$_{0.8}$Sr$_{0.2}$NiO$_2$ thin films have bulk superconductivity instead of interface effect. Another two systems, namely hole doped LaNiO$_2$ and PrNiO$_2$ were synthesized successfully and found to be superconductive as well.[51-54] Accordingly, the nickelate system is considered to be a promising candidate to be expanded into a large family, providing a new route to research and explore HTS. Based on the current experimental results, one can see that these three systems La, Pr, Nd 112 phases share similar electronic structures and superconducting properties. Thus in the following sections of this review, we mainly focus on discussing about the representative NdNiO$_2$ system, including the calculated electronic structures with multiband features, spin configurations, magnetic excitations, phase diagrams as well as the measurement of single particle tunneling spectrum, finally we make some open discussions for the possible pairing symmetries and pairing mechanism in the newly found Ni based superconductors.

## ELECTRONIC STRUCTURES

As we know, the parent compounds of cuprates may be described as a Mott insulator with AF long-range order[55-59] and superconductivity occurs upon chemical doping.[60] Due to the strong *p-d* orbitals hybridization,[61] the doped holes enter into the oxygen sites in the $CuO_2$ planes. They are proposed to combine with the $3d_{x^2-y^2}$ spins of Cu ions to form the Zhang-Rice singlets,[62] moving through the square lattice and exchange with their neighboring Cu spins. This theoretical proposal sets cuprates into a single band limit and naturally leads to an effective two-dimensional *t-J* model to describe the low-energy physics of cuprates. In FeSC the situation exhibits more complexity because the As/Se locates alternatively above and below the center of the Fe square lattice. The crystalline environment experienced by Fe atoms is somewhat in between a tetrahedral one, in which the energy of the $t_{2g}$ orbitals is higher than that of the $e_g$ orbitals, while for an octahedral one as in cuprates, the energy of the $t_{2g}$ orbitals is lower. As a result, the crystal splitting between the orbitals is weakened and all five *d*-orbitals give a considerable contribution to the low-lying energy of electronic states.[63] As for the newly found Ni based superconductors, many papers have made detailed theoretical calculations of the electronic structures, demonstrating the multi-orbital features in Nd/La 112 system within the low energy region and pointing out both similarities and differences between nickelates and cuprates.[64-75]

**1. Mott Insulator, Self-Doping Effect and Possible Kondo Coupling**

When we compare $NiO_2$ layer in nickelate and $CuO_2$ layer in cuprates, a major difference becomes immediately obvious. As shown by the sketch in Figure 2A, a charge-transfer energy is estimated to be $\Delta \approx 9$ eV in $NiO_2$ and $\Delta \approx 3$ eV in $CuO_2$. According to the Zaanen-Sawatzky-Allen (ZSA) scheme,[76] cuprates locate in the regime of charge transfer insulator while nickelates belong to the system of Mott insulator.[77] The holes doped in a Mott insulating $NiO_2$ layer would reside on the Ni-derived bands, not in the O $2p$ one. Because $Ni^{2+}$ ($3d^8$) with $S = 1$ is very common in all the known $Ni^{2+}$ oxides, this makes the appearance of rather high-$T_c$ superconductivity very puzzling and definitely unlike that in cuprates. In experiments, Hepting et al.[78] conduct X-ray absorption spectroscopy (XAS) and emission spectroscopy (XES) measurements near the O K-edge in parent phase $R$NiO$_2$ ($R$ = La, Nd), roughly reflecting the occupied and unoccupied oxygen partial density of state (PDOS), respectively. As shown in Figure 2B, the oxygen PDOS exhibits a diminished weight near the Fermi energy, especially in the unoccupied states, also indicating that O $2p$ orbitals carry less weight in the expected upper Hubbard band by comparison. All these are consistent with the calculated oxygen PDOS from the calculation of LDA + $U$ method (inset). The inset also plots the sketch of the relationship between $U$ and $\Delta$, showing the significant distinction to cuprates. Additionally, Figure 2C shows the results of resonant inelastic X-ray scattering (RIXS) at the Ni L$_3$-edge (a core-level 2$p$ to valence 3$d$ transition). The markers A indicate the main absorption peak of LaNiO$_2$ and NdNiO$_2$ which resembles the single peak associated with the $2p^63d^9$–$2p^53d^{10}$ transition in cuprates.[79] The A' labels highlight the hybridization between the Ni 3$d_{x^2-y^2}$ and $R$ 5$d$ orbitals. In this configuration, the Ni state can have a charge transfer to the rare-earth cation, thus leaving holes in Ni orbitals and electrons in $R$ 5d orbitals, this is the so-called self-doping effect. In NdNiO$_2$, the similar feature due to the Nd–Ni hybridization also exists in RIXS, but its resonance energy A' almost coincides with the main peak A. Based on above experimental results, one can naturally understand that in parent phase the single occupied Ni 3$d_{x^2-y^2}$ orbital with strong correlation may give rise to a local spin 1/2 and a Mott insulator state with AF long-range order. However, the parent compound NdNiO$_2$ displays metallic behavior at high temperatures (see Figure 1B), and shows no sign of any magnetic long-range order in the whole measured temperature range.[37,38] Similar results have also been found previously in LaNiO$_2$.[80] These experimental observations show an obvious contradiction to the naive expectations. It is therefore important to address what is the nature of the parent compounds and how the AF long-range order is suppressed. Zhang et al.[81] make a detailed analysis of resistivity data as functions of temperature for both parent compounds NdNiO$_2$ and LaNiO$_2$. When the data are put on a semi-logarithmic scale, they find that the resistivity $\rho$ upturn well obeys a logarithmic temperature (ln $T$) dependence below about 40 K down to 4 K for NdNiO$_2$ and below about 70 K down to 11 K for LaNiO$_2$. This is supposed to be the evidence of magnetic Kondo scattering, which is further supported by the Hall effect data with the same ln $T$ dependence at low temperatures.[46,80] Considering the existence of self-doping effect and the more itinerant $R$ 5$d$ electrons, a schematic picture is plotted in Figure 2D. The presence of both the Kondo singlets and the holons can suppress very efficiently the AF long-range order and cause a phase transition from the Mott insulating state to a metallic state. However, we would like to remind that, the correlation effect will also induce a low temperature up-turn of resistivity, as that occurring in under doped cuprates.[82,83] Thus it remains interesting to unravel whether the Kondo scattering mechanism is the dominant one to interpret this low-temperature upturning of resistivity.

2.  **Two Orbitals, Three orbitals or More**

For modeling the realistic electronic structure in Nd/La 112 system, the first question appears to be how many orbitals should be considered to capture the basic features of both the normal state and superconducting state. Kitatani et al.[84] propose that $R$NiO$_2$ can be described by the one band Hubbard model, albeit with an additional electron reservoir, which is used to calculate the critical temperature $T_c$. Hepting et al.[78] construct a two-orbital model of Ni-$d_{x^2-y^2}$ orbital and a $R$-$d_{z^2}$ orbital to study the hybridization effects between them. Some people proposed different types of two-orbital models consisting of two Ni-$d$ orbitals. Hu et al.[85] include Ni-$d_{x^2-y^2}$ and Ni-$d_{xy}$ orbitals, while Zhang et al.,[86] Werner et al.,[87] and Wan et al.[88] include Ni-$d_{x^2-y^2}$ and Ni-

$d_z^2$ orbitals. These types of two-orbital models are aimed to study the competition of high spin $S = 1$ doublon and low spin $S = 0$ singlet when the system is hole doped.[89] Some people construct three-orbital models. Wu et al.[90] include Ni-$d_{x^2-y^2}$, Nd-$d_{xy}$, and Nd-$d_z^2$ orbitals. This model is further used to calculate the spin susceptibility and robust $d$-wave pairing symmetry in Nd$_{1-x}$Sr$_x$NiO$_2$. Nomura et al.[91] include Ni-$d_{x^2-y^2}$ orbital, R-$d_z^2$ orbital, and a bonding orbital made from interstitial-$s$ orbital in the Nd layer and the Nd 5$d_{xy}$ orbital, which is used to study the metallic screening effects of the Nd-layer states on the Hubbard $U$ between Ni-$d_{x^2-y^2}$ electrons. Gao et al.[92] construct a general four-orbital model which consists of two Ni-$d$ orbitals and two R-$d$ orbitals. The model is used to calculate the topological property of the Fermi surface. Jiang et al.[93] use a tight-binding model that consists of five Ni-$d$ orbitals and five R-$d$ orbitals to comprehensively study the hybridization effects between Ni-$d$ and R-$d$ orbitals. Jiang et al. also highlight the $f$ orbital of Nd hybridizes with Ni-$d_{x^2-y^2}$, which is a non-negligible ingredient for transport and even high-temperature superconductivity. Botana et al.,[94] Lechermann,[95] and Karp et al.[96] consider more orbitals (including Nd-$d$, Ni-$d$, and O-$p$ states) in the modeling of NdNiO$_2$ with the interaction applied to Ni-$d$ orbitals and make a comparison to infinite-layer cuprates. Botana et al.[94] extract longer-range hopping parameters and the $e_g$ energy splitting. Lechermann[95] studies hybridization and doping effects. Karp et al.[96] calculate the phase diagram and estimate the magnetic transition temperature.

Sakakibara et al.[97] made the first-principles calculation in LaNiO$_2$, modeled by seven orbitals constituted of five Ni 3$d$ orbitals (3$d_{x^2-y^2}$, 3$d_z^2$, 3$d_{xz}$, 3$d_{yz}$, 3$d_{xy}$) and two

La 5$d$ orbitals (5$d_{xy}$, 5$d_{z^2}$). Figure 3A shows the first-principles band structure of LaNiO$_2$, superposed by Wannier-orbital weight. The corresponding Fermi surfaces at $k_z$ = 0 and $k_z$ = π planes are also plotted both for parent phase (solid lines) and 20% hole doping (dashed lines). The main Fermi surface in the system named by α is constructed from Ni 3$d_{x^2-y^2}$ orbital which plays the dominant role in the low-lying energy of the system. It displays a van Hove feature evolving from the $k_z$ = 0 cut to the $k_z$ = π cut with the evolution from a hole pocket around the $M$ point in the $k_z$ = 0 plane to an electron pocket around the $Z$ point in the $k_z$ = 0 plane. One can also see two small electronlike pockets β and γ around the $\Gamma$ and $A$ points which are contributed mainly from La-derived orbitals. These self-doping bands are crucially originated from the hybridization between Ni 3$d$ and La 5$d$ orbitals. More specifically, β pocket around the $\Gamma$ point has the mixture of La 5$d_{z^2}$ and Ni 3$d_{z^2}$ while γ pocket around the $A$ point has the mixture of La 5$d_{xy}$ and Ni 3$d_{xz/yz}$. The similar band structure of NdNiO$_2$ can be obtained in density functional theory (DFT),[95] as shown in Figure 3B. With 20% hole doping, both β and γ pockets around the $\Gamma$ and $A$ points will diminish. However, the β pocket in LaNiO$_2$ disappears almost completely and it still exists in NdNiO$_2$, leaving a smaller γ pocket around the $A$ point in both compounds.

To conclude, for the mother compound, a small amount of holes are self-doped into the Ni 3$d_{x^2-y^2}$ orbital due to the presence of electron pockets. These electron pockets may have relevance to the metallic behavior observed experimentally, while the electronic state of the $d_{x^2-y^2}$ band with a hole self doping may be close to that of the heavily underdoped cuprates with no magnetism or superconductivity.[98-102]

Furthermore, the superconductivity in thin films of $Nd_{1-x}Sr_xNiO_2$ may share some similarities like that of the single-orbital cuprates, while more distinctions are expectable, providing a new fascinating platform on the research of unconventional correlated superconducting materials.

3. **Transport Measurements and Phase Diagram**

Considering the difficulty in synthesizing the high quality superconducting nickelate thin films, we just make a brief review on the transport data and the resultant phase diagram. Li et al.[103] have made a detailed study on resistivity and Hall effect measurements on the typical system of $Nd_{1-x}Sr_xNiO_2$ infinite layers. Figure 4A displays the temperature-dependent in-plane resistivity $\rho_{xx}$ across the doping series. The films with x = 0.15, 0.175, 0.2, and 0.225 show varying $T_c$, while for x = 0, 0.1, 0.125, and 0.25, a weakly insulating phase emerges at low temperatures, as shown in the upper panel of Figure 4B. These data indicate a superconducting dome that is similar to hole-doped cuprates[104,105] in the lower panel, but approximately half as wide in the doping regime. Furthermore, an approximately linear $T$ dependence of $\rho_{xx}$ is observed across a wide temperature range above $T_c$, similar to that found in cuprates (known as the "strange metal" phase[106-108]) and other strongly correlated systems, suggesting a similar possible origin for $\rho_{xx}(T)$ despite different underlying electronic structures.[109] As for the noticeable upturn in $\rho_{xx}(T)$ for nonsuperconducting compositions (x = 0, 0.1, 0.125, and 0.25), it is not simply identified as weak localization. Also, the lack of strongly insulating behavior has already been widely reported for the undoped compounds.[46] It can be possibly attributed to the self-doping bands and Kondo effect. They further find that the "overdoped" regime does not appear to approach the Fermi liquid ending point commonly understood in the hole-doped cuprates.[110-113] Overall, the relevance of "hole doping" remains an open question for $Nd_{1-x}Sr_xNiO_2$. They also measure the evolution of the normal state Hall effect on this series of samples. Figure 4C shows that $R_H$(20 K) monotonically increases from negative to positive values, crossing zero between x = 0.175 and 0.2. While in $La_{2-x}Sr_xCuO_4$, $R_H$ was found to be large and positive in the undoped case, and systematically varied as ~1/x with initial doping.[104,114] The distinctive doping dependence of $R_H$ in the nickelates is clearly inconsistent with single-band hole doping. A simple explanation can be considered is a two-band model with both electron-like and hole-like Fermi surfaces. With increasing hole doping, a predominantly electron-like Hall effect will undergo a transition to a hole-like one, followed by the reduced Fermi level. This two-band picture is well corroborated by many recent electronic structure calculations, which are discussed in detail in the former section. The general consensus is that band structures for $NdNiO_2$ is the presence of a large hole pocket with Ni $3d_{x^2-y^2}$ character near the $k_z$=0 cut, and two electron pockets with the dominant characters of Nd $5d_{xy}$ and $5d_{z^2}$ orbitals.

Additionally, it can be noted that bulk samples with the same Nd-112 structure and proper Sr-doping as well as $Sm_{0.8}Sr_{0.2}NiO_2$ have been made by some of us,[115,116] but no superconductivity is observed, as shown in Figure 4D. The absence of superconductivity in bulk samples has recently been repeated by another group.[117] In addition, the bulk samples of Nd1-xSr$_x$NiO exhibit very strong insulator-like behavior even under a pressure up to 50.2 GPa. It was argued that the absence of superconductivity in bulk samples may be induced by the heavy deficiency of Ni, or due to the significant influence of physical properties by the subtle change of the structure and doping level between the bulk and film samples. The Ni deficiency can either strongly lower down the hole doping or act as strong scattering centers which localize the mobile electrons and thus diminish superconductivity. It seems to be physically unreasonable of attributing the absence of superconductivity in bulk samples to intercalating extra hydrogen,[118] since both the superconducting thin films and the bulk samples are treated with CaH$_2$ in exactly the same way. Furthermore, we also successfully synthesize 112 phase using a non-hydrogen reducing medium zirconium, and find that although the 112 phase has a better crystallinity, the insulating behavior seems to be even stronger, and superconductivity remains absent. Despite a lot of challenge, it is still worth to have more efforts to figure out the unsettled issue.

## SPIN CONFIGURATIONS AND MAGNETIC EXCITATIONS

As we know, the parent phase of cuprates is depicted as a charge transfer insulator with an AF order,[119-122] where the Cu$^{2+}$ ions have one active $d_{x^2-y^2}$ orbital hybridized with the $p$ orbitals of the neighboring in-plane oxygens.[123,124] Upon hole doping, due to the negative charge-transfer gap, the holes predominantly go to the O-$p$ orbitals. Thus it will give rise to the spin configuration of 3$d^9$ Cu$^{2+}$, surrounding by a ligand hole on oxygen. This is the formation of Zhang-Rice singlets.[62] In inelastic neuron scattering measurements, a square-shaped continuum of excitations peaked at incommensurate positions, as well as an 'hour glass' shape of the magnetic dispersions in the superconducting state of cuprates has been extensively observed.[125-128] These excitations are regarded as a general property of cuprates, and a promising candidate for magnetic mediated electron pairing. Moreover, when revisiting the phase diagrams of temperature versus doping level both in cuprates and FeSC, we can find that the occurrence of superconductivity is intimately correlated with the disappearance of AF long-range order and the AF fluctuations.[129] As a result, it may lead to the unconventional pairing state of $d$ wave in cuprates[130] and $s^{\pm}$ in FeSC.[131] Accordingly, it is highly desirable to address the spin configurations and magnetic excitations in nickle based superconductors.

**1. Competition Between High Spin and Low Spin States**

In nickelates, the charge-transfer energy is much larger so one may expect $Ni^{2+}$ to have the dominant holes residing on the Ni site, rather than in the O-$p$ band. As depicted in Figure 5A, in a square lattice environment, the $e_g$ states may be energetically arranged in two different ways, which depends on the limit whether the Hund's rule coupling is larger than the crystal-field splitting between the two $e_g$ orbitals. For high spin (HS) state, the electron from the $d_{z^2}$ orbital is removed, in which case the resultant configuration is $(t_{2g})^6 (d_{z^2})^1 (d_{x^2-y^2})^1$ with $S = 1$; For low spin (LS) state, the electron from the $d_{x^2-y^2}$ orbital is removed, in which case the configuration is $(t_{2g})^6$

$(d_z^2)^2$ with $S = 0$. Concerning that the $Ni^{2+}$ ions are commonly found to be spin-triplet state in most of the nickel compounds, it has recently been argued in several theoretical works[132,133] that hole-doping $Nd_{1-x}Sr_xNiO_2$ should produce $Ni^{2+}$ with spin $S = 1$. Based on the HS state, Zhang et al.[86] propose a variant of the *t-J* model and find two distinct mechanisms for *d*-wave superconductivity. However, Jiang et al.[134] first point out that the $S = 1$ state may be incompatible with robust superconductivity, and a number of many-body calculations using LDA and dynamical mean field theory (DMFT) pointed to the formation of intraorbital singlets.[95,96] Krishna et al.[135] find that from the first-principles calculations of explicit Sr doping in $R$NiO$_2$ ($R$ = La, Nd) supercells, a LS state is favored. From experimental measurements, Rossi et al.[136] use a combination of high-resolution XAS and RIXS and find that doped holes are primarily introduced in the Ni $3d_{x^2-y^2}$ Hubbard band in a low-spin configuration. Wan et al.[88] discuss the solutions of an effective two-band model including Ni-$3d_{x^2-y^2}$/Ni-$3d_{z^2}$ orbitals on the basis of dynamical mean field theory. As shown in Figure 5B, whether an $S = 0$ or 1 state emerges depends on a precise value of the intra-atomic Hund's coupling $J_H$ in the vicinity of its commonly accepted range of values 0.5-1 eV. Until now, we cannot make a definite conclusion about whether the $S = 0$ or 1 scenario is realized for doped nickelates. The key issue is how to extract the real strength of Hund's coupling in the system. Therefore, the spin state of $Ni^{2+}$ in hole-doping $Nd_{1-x}Sr_xNiO_2$ remains highly debatable and worthy of more attention.

## 2. AF Magnetic Excitation

Most intriguingly, one key experimental observation for the infinite-layer NdNiO$_2$ is that its resistivity exhibits a minimum at about 70 K and an upturn at a lower temperature. At the same time the Hall coefficient drops towards a large value, signaling the loss of charge carriers. More interestingly, no long range magnetic order has been observed in powder neutron diffraction on LaNiO$_2$ and NdNiO$_2$ when temperature is lowered down to 5 K and 1.7 K, respectively.[37,38,80] This greatly challenges the existing theories, since it is generally believed that magnetism is essential for the emergence of unconventional superconductivity. Therefore, it is highly desirable to study the magnetic properties of undoped parent NdNiO$_2$ and elucidate its experimental indications.

From theoretical aspects, Gu et al.[137] find that the hybridization between Ni-$d_{x^2-y^2}$ orbital and itinerant electrons in $R$NiO$_2$ is substantially stronger than previously thought. Because of that, Ni local moment is screened by itinerant electrons and the critical $U_{Ni}$ for long-range magnetic ordering is increased. Liu et al.[138] present a first-principles calculation for the electronic and magnetic structure of undoped parent NdNiO$_2$. By taking relevant interaction strength found experimentally into account, they found that the antiferromagnetic with ($\pi$, $\pi$, $\pi$) vector in NdNiO$_2$ is a compensated bad metal with small Fermi pockets. In Figure 5C, the white ball represents Ni site with local moment and the four exchange coupling parameters are indicated by the blue arrows. The calculated exchange coupling parameters as a function of Hubbard $U$ are shown in Figure 5D. One can get that the estimated magnetic exchange interaction is around $J_1 \sim 10$ meV, $J_2 \sim 10$ meV, $J_3 \sim -13$ meV, $J_4 \sim 0$ meV. This indicates that effective exchange interactions in NdNiO$_2$ are about one-order smaller than those of cuprates ~112 meV.[139-141] Also it results in a relative weaker magnetic ordering and lower Neel temperature $T_N$ in NdNiO$_2$, compared with cuprates. Furthermore, the infinite-layer nickelate is believed to be a worse metal compared to elemental nickel with Hubbard $U$ about 3 eV,[142] which can be considered as a lower boundary of $U$. The Coulomb interaction in infinite-layer nickelates should be smaller than that in the charge-transfer insulator NiO with $U$ about 8 eV,[143] which can be considered as an upper boundary of $U$. Therefore, a reasonable value of $U$ in NdNiO$_2$ can be estimated around 5-6 eV. As shown in Figure 5E, there could exist a transition from normal metal to bad AFM metal around $T_N \sim$ 70-90 K, which provides a plausible understanding of minimum of resistivity and Hall coefficient drop in infinite-layer NdNiO$_2$.

From experimental aspects, Cui et al.[144] report the $^1$H NMR measurements on powdered Nd$_{0.85}$Sr$_{0.15}$NiO$_2$ samples by taking advantage of the enriched proton concentration after hydrogen annealing. The spin-lattice relaxation rate $T_1^{-1}$ is a sensitive probe of low-energy spin fluctuations.[145,146] The temperatures dependent $^1T_1^{-1}$ under various fields is plotted in Figure 6A. From 2 K, the $^1T_1^{-1}$ first increases dramatically, then forms a peaked feature at $T \sim$ 40 K, finally flattens out at temperatures above 100 K. This peak feature is also demonstrated by the spin-recovery curve for the $T_1$ as shown in the inset of Figure 6A, where the spin recovers more rapidly at 40 K than that at 200 K and 10 K. A sharp peak in $^1T_1^{-1}$ is usually a signature of magnetic phase transition, while the broad one at about 40 K suggests the onset of a short-range glassy anti-magnetic order in bulk $Nd_{0.85}Sr_{0.15}NiO_2$, which is similar to underdoped cuprate superconductor.[147] Furthermore, the plot of a log-log scale is shown in Figure 6B. Below 40 K, $^1T_1^{-1} \propto T^\alpha$, following a power-law behavior with a low power-law exponent $\alpha$ = 2, much smaller than that caused by AF spin wave excitations ($\alpha$ = 5).[148] This indicates the onset of low-energy spin fluctuations and remaining spin excitations extending to much higher temperatures. The finding of strong AF fluctuations reveals the strong electron correlations in bulk $Nd_{0.85}Sr_{0.15}NiO_2$ and pave the way for understanding the relationship between magnetism and superconductivity in nickelates. Lu et al.[149] measure the dispersion of magnetic excitations in undoped $NdNiO_2$ by using RIXS at the Ni $L_3$-edge. Figure 6C plots a summary of fitted magnetic mode energy $\varepsilon$ (filled red circles) and damping factor $\gamma$ (empty red circles) versus in-plane momentum transfers $q_{//}$ along high-symmetry directions. $NdNiO_2$ possesses a branch of dispersive excitations with a bandwidth of approximately 200 meV, which can be fitted to the linear spin wave theory of a two-dimensional antiferromagnetic Heisenberg model. The significant damping and rather constant $\gamma_q$ of these modes indicates the rare-earth itinerant electrons play a role here, which are highly coupled to Ni-3$d$ orbitals.

To reconcile the discrepancy between the observed paramagnetic metallic state (only the remaining AF excitations) of $RNiO_2$ and theoretically predicted AF transition nearby 90 K, one should be noticed that the estimation of Neel temperature in Figure 5E does not take the influence of conduction electrons into account. Considering the strong coupling between Ni-$d_{x^2-y^2}$ and itinerant electrons originated from Nd 5$d$ orbitals, it will screen the Ni local moment as in Kondo systems and increase the critical value of $U_{Ni}$ that is necessary to induce long-range magnetic ordering. As a result, the local magnetic AF ordering below the Neel temperature should be very weak, which may be hard to be visualized in the magnetic related measurements, such as NMR and neutron scattering. Instead, only the remaining AF fluctuations can be detected.

## SUPERCONDUCTING STATE

The foregoing sections are mainly focused on the electronic structures and magnetic excitations of normal state in $R$NiO$_2$, now we will turn to discussing about the superconducting properties of this newly discovered superconducting nickelate Nd$_{1-x}$Sr$_x$NiO$_2$ thin film, including upper critical field, superconducting gap and possible pairing symmetries.

1. **Isotropic Pauli-Limited Pairing State**

As we know, in most superconductors,[150] the superconducting gap ($\Delta$) and the pairing strength may be linked to the upper critical field $H_{c2}$ via the Pippard relation $\xi = \hbar v_F/\pi\Delta$ and $\mu_0 H_{c2} = \Phi_0/2\pi\xi^2$ (orbital depairing). Here, $\xi$ is the coherence length, $v_F$ is the Fermi velocity, and $\Phi_0$ is the flux quantum. However, in a few superconductors, Cooper pairs can be broken mainly due to the Zeeman-split effect,[151] and thus the upper critical field is dominated by Pauli paramagnetic limit $\mu_0 H_P = \sqrt{2}\Delta/g\mu_B$. Here $\mu_B$ is the Bohr magneton, and $g$ is the Landé factor. Therefore, it is worthy measuring the upper critical field to obtain the information of the pairing strength for this new superconducting system. More generally, $Nd_{1-x}Sr_xNiO_2$ system can be regarded as layered superconductor, in which case one may naturally wonder whether there is some anisotropy in $\Gamma = H_{c2//}/H_{c2\perp}$, where $H_{c2//}$ and $H_{c2\perp}$ represent the upper critical fields when it is along the ab plane and the crystal c-axis direction, respectively. The case for the nickelates is a priori interesting: although it shares the same crystal structure as infinite-layer cuprates, calculations indicate that both a quasi-2D hole band and three-dimensional electron bands are present in the nickelates for a broad range of electron interactions.[64,94] The relative importance of these bands is a subject of much interest with respect to the superconducting state. Moreover, the reported $Nd_{1-x}Sr_xNiO_2$ thin film can be synthesized as thin as only about 10 nm,[152] thus providing a new system to explore longstanding debates on the role of dimensionality for superconductivity.

Xiang et al.[153] extract $H_{c2//}(T)$ and $H_{c2\perp}(T)$ from $\rho(T)$ curves under different magnetic fields on $Nd_{0.8}Sr_{0.2}NiO_2$ thin films, by using two different criteria of 95% $\rho_n(T)$ and 98% $\rho_n(T)$, where $\rho_n(T)$ is the linear extrapolation of the normal-state resistivity. Since the normal-state residual resistivity is large for this film, they attempt to use the Werthamer, Helfand and Hohenberg (WHH) theory[151] in a dirty limit for a superconductor with a single s-wave gap to fit the data, as shown in Figure 7A. They give several innovative experimental facts, (i) the Maki parameter α ranges from 13 to 42, when the magnetic field is applied within ab plane or along c-axis and under different criteria. The obtained very large α here is very rare, even much larger than that in the most well-known Pauli limited systems, such as heavy-fermion and organic superconductors.[154-160] Meanwhile, a very large value of α indicates the possible existence of the Fulde–Ferrell–Larkin–Ovchinnikov (FFLO) state in the high-magnetic-field region at low temperatures.[161-163] Since the FFLO state is fragile in the presence of disorders,[157,164] the existence of this state in the $Nd_{1-x}Sr_xNiO_2$ thin film requires further investigation via high-magnetic-field experiments; (ii) the value of $\Gamma$ approaches to 1 when the temperature is well below $T_c$, namely the very small anisotropy between $\xi_{ab}$ and $\xi_c$ (equivalent to the effective mass ratio of $m_c^{1/2}/m_{ab}^{1/2}$). This provides strong evidence that the existence of two 3D electron pockets makes the electronic structures more isotropic, when compared to the only one single band of quasi 2D Ni $3d_{x^2-y^2}$; (iii) $H_0(T)$ values are determined using the criteria 0.1% $\rho_n(T)$ and 1% $\rho_n(T)$, in order to extract the irreversibility field of the thin film.

Wang et al.[165] use similar measurements to probe the anisotropy in $Nd_{0.775}Sr_{0.225}NiO_2$ with a slightly lower $T_c$. Figure 7B plots superconducting $H_{c2}$-$T$ phase diagrams for magnetic fields along the c-axis and in the ab plane, including many intriguing phenomena. (i) The key observation is a $T$-linear dependence of $H_{c2\perp}$ and a $(T_c - T)^{1/2}$ dependence of $H_{c2//}$. First of all, they rule out the possibility of 2D superconductivity. The main reason is that they deduce a thickness to be 2.3-3 times larger than the observed one, if adopting the linearized Ginzburg–Landau model of 2D superconductor.[166] Another plausible origin is that a Pauli limited superconductor can also give a $(T_c - T)^{1/2}$ dependence of $H_{c2//}$;[167-169] (ii) high values of Maki parameter α for both orientations, which indicates the strong presence of the paramagnetic pairing breaking effect; (iii) high values of α suggests possible existence of FFLO state at low temperatures and high magnetic fields, which seems to be consistent with the upturn of $H_{c2}$ data below 4 K. However, the fact that the film lies in the dirty limit is inconsistent with anticipated expectations. The upturn of $H_{c2}$(T) could indicate the occurrence of two-band superconductivity.[170]

To draw a conclusion about the magnetotransport measurements of the two works above, they both point out the isotropic Pauli limited superconductivity with unconventional pairing in the newly found nickelate superconducting film. This provides strong evidence that two small 3D electronlike pockets β and γ contributed mainly from Nd-derived orbitals play a non-negligible role here. In addition, a very large value of Maki parameter α indicates possible existence of the very interesting FFLO state in the high-magnetic-field region at low temperatures under the condition that the system can be pushed into a clean limit.[157]

2. **Single Particle Tunneling Spectrum Measurement**

Concerning the pairing mechanism of nickel based 112 systems, the core issue is to know the superconducting gap function which measures the pairing interaction of the two electrons of a Cooper pair. However, the research is very rare on physical properties of this material, especially for APERS and STM measurements. Therefore, it is urgent to conduct spectroscopic measurements to directly determine the gap structure. In this section, we introduce the first investigation of single particle tunneling measurements on the superconducting $Nd_{1-x}Sr_xNiO_2$ thin films.[171] Figure 8A shows the surface of the film just after annealing by the soft-chemistry method. One can see that the surface is not atomically flat showing a roughness of about 1~2 nm. This large roughness may be induced by a drastic reaction of the 113 film with hydrogen during the post-annealing process. However, if we take a long time vacuum annealing (at about 180 °C in a vacuum of $10^{-9}$ torr for 12 h) on the film with this type of rough surface, some areas of the surface show layer-by-layer structure with terraces, a typical surface morphology is shown in Figure 8D. Now the roughness becomes much smaller, in which case the tip can have a better stabilization during the tunneling process. We have conducted measurements of scanning tunneling spectroscopy (STS) on the surfaces with these two different morphologies, one is called as rough surface, another one is called as smoother surface. We find that the superconducting spectra predominantly show two types of features on the rough surfaces. One type shows a typical V-shape feature, which is shown in Figure 8B. By doing the Dynes model fitting,[172,173] as displayed by the red curve, we find that the spectrum can be nicely fitted with a d-wave gap of $\Delta = 3.9\cos2\theta$ (meV). The other one shows a full gap feature plotted in Figure 8C, with a Dynes fitting of $\Delta = 2.35(0.85+0.15\cos4\theta)$ (meV). A slight anisotropy (about 15% weight of the differential conductivity) is added to the gap function in order to have a good fit. This may suggest that at least one of the bands is fully gapped. On the smoother surface, we find dominant V-shape spectra as shown in Figure 8E. It needs to be emphasized that, this type of full gap is hardly observed on the smoother surface. Instead, we can easily observe a mixture of the two-gap features on the spectra, as shown in Figure 8F. The Dynes fitting is $\Delta_1 = 5.3(0.8\cos2\theta+0.2\cos6\theta)$ (meV), taking a dominant weight of $p_1 = 85\%$ and $\Delta_2 = 2$ meV.

Furthermore, one can note that the topographic image in Figure 8A shows strong roughness, which provides the possibility for the STM tip to detect tunneling behavior along different directions at different positions. This may give us the advantage to detect the superconducting gap features derived from different bands, which could be the reason for us to see two distinct gap structures at different positions.[174,175] The same situation occurs in the STM measurements of $MgB_2$ bulk and film.[176] The STM tip can detect the gap with a magnitude of about 7.1 meV on the σ-band on some grains, and can also measure the gap on the π-band with the value of 2.3 meV on other grains. However, as mentioned above, on the smoother surface, it is hard to observe a "clean" full gap feature. Most time the spectrum shows a mixture of the two and a robust V-shape feature appears near zero bias. This may be understood in the way that, now the tunneling current mainly goes along c-axis direction, with a reduced side tunneling component which would occur in the measurements on the rough surface at some locations. Based on above discussion and the multiband features in $Nd_{1-x}Sr_xNiO_2$, one may naturally conclude that the two kind of features on the measured spectra correspond to the gaps on different Fermi surfaces.

**3. Robust *D*-Wave Pairing and Possible Explanations for Full Gap**

Considering the common feature of V-shape spectra measured on superconducting thin films, we first discuss about the origin of the *d*-wave gap. It has been extensively demonstrated in many theoretical papers that $Nd_{1-x}Sr_xNiO_2$ system has a robust *d*-wave pairing, in both limits using weak-coupling approach of random phase approximation (RPA) and strong coupling of *t-J* model. Wu et al.[90] find that, in all instances, the dominant pairing tendency is in the $d_{x^2-y^2}$ channel. In analogy with cuprates,[12,130] pnictides,[27,131] and heavy-fermion superconductors,[177,178] superconductivity in the present compound is assumed to be mediated by spin-fluctuation. Based on the multiorbital fluctuation exchange approximation, the effective pair interaction vertex $\Gamma(k,k')$ is determined by RPA susceptibility. By solving the dimensionless pairing strength functional, the largest pairing eigenvalue $\lambda$ will lead to the highest transition temperature and its eigenfunction determines the symmetry of the gap.[179-181] Figure 9A displays the bare susceptibilities for n = 0.8 filling. The dominant peaks locate around the *M* and *A* points, indicating intrinsic antiferromagnetic fluctuations. The prominent features in the orbital-resolved susceptibility are that the peaks around *M* and *A* are mainly attributed to the Ni $d_{x^2-y^2}$ orbital, which will further promote the dominant $d_{x^2-y^2}$-wave pairing state. As the effective low-energy interaction parameters remain undetermined for $Nd_{1-x}Sr_xNiO_2$, a large region of parameter space has been calculated within RPA calculations. The obtained pairing eigenvalues as a function of interaction strength $U_{Ni}$ for n = 0.8 are displayed in Figure 9B. One can find that the $d_{x^2-y^2}$ pairing state is dominant over other competing ones, for example for $d_{xy}$ state. This is consistent with the fact that both the dominant density of states and pairing interactions reside on the Ni $d_{x^2-y^2}$ orbital.

The nickelate system seems to belong to the intermediate coupling one, it thus becomes important to resolve the pairing mechanism. From a strong-coupling perspective, concerning the nickelate system, the *t-J* model can be reduced to one single Ni- $d_{x^2-y^2}$ orbital for simplicity. Wu et al.[90] investigate the pairing state for an extended range of doping levels. Figure 9C plots the representative superconducting gap of the $d_{x^2-y^2}$ pairing versus different dopings. It shows that there is a superconducting dome and the gap reaches the maximum at a doping of 0.1 hole/Ni. Moreover, the 3D gap function of the obtained $d_{x^2-y^2}$-wave pairing is displayed in Figure 9D at 0.2 hole doping, without considering the contribution of Nd orbitals. The main findings from a $t$-$J$ model analysis are consistent with the RPA analysis.

Now it can be understood that the measured V-shape spectra featured a $d$-wave gap is mainly originated from the Ni-$3d_{x^2-y^2}$ orbital, however, several theoretical assumptions are proposed to explain the observed full gap. The first picture is that, this full gap may just simply reflect the gap function on the hybridized orbitals of the Ni $3d_{z^2}$, $3d_{xz,yz}$ mixed with the Nd $5d_{z^2}$, $5d_{xy}$, namely on the β and γ Fermi pockets. However, if we just simply follow the $d_{x^2-y^2}$ notation for the gaps in the whole momentum space, the nodal line will cut the two electron pockets, which is not consistent with our observation of a full gap. Adhikary et al.[182] pointed out the orbital-selective superconductivity in a two-band model of infinite-layer nickelates. At the basis of the proposal, we can plot a cartoon picture for the Fermi surfaces and the gap structure on different cuts of $k_z$, as depicted in Figure 10. Concerning the inter-orbital Hubbard interaction between different orbitals, we expect not only a $d$-wave gap on the α pocket, but also full gaps on the β and γ ones with opposite gap signs. The latter is a bit like the $s^{\pm}$ pairing in many pnictides. This scenario is quite interesting, and tells that not only the intra-pocket interaction, but also the inter-pocket interaction plays an important role here, leading to the orbital selective pairing. The second one to interpret this full gap relies on a recent theoretical calculation about the doping dependent phase evolutions of the pairing symmetries.[183] This phase diagram was established based on the proposed picture of self-doped Mott insulator,[81] which shows an evolution from a *d*-wave dominant region to an *s*-wave region with the intermediate phase of *d*+i*s* wave. To satisfy this model, we need to postulate that the doping level is not homogeneous in the film, thus somewhere the system shows a *d*-wave gap, somewhere *s*-wave and somewhere a mixture of the two. This seems compatible with our results, however, according to our experience of MBE growth, the doping level of Sr may not vary too much in the deposition region (5×5 mm$^2$) of the thin film, but rather the local clustering or reconstruction may give more influential effects. The third one suggests that the full gap of the tunneling spectrum may arise from the NiO$_2$ terminated surface which has natural buckling planes of NiO$_2$.[184] This picture is also interesting, which can be checked out on a surface with atomically resolved morphology. The fourth one proposes that the experimental observation can be simply explained within a pairing scenario characterized by a $d_{x^2-y^2}$-wave gap structure with lowest harmonic on the Ni-band and higher-harmonics on the Nd-band.[185] This scenario can be tested in future STM experiments with improved sample quality where the position of STM tip with respect to the Nd$_{1-x}$Sr$_x$NiO$_2$ unit cell can be precisely determined. Last but not the least, the full gap feature may be just induced by the tunneling matrix problem. The symmetry of superconductor gap function refers to its transformation under crystal symmetry operations. In this general point of view, even a *p*-wave or a time-reversal-symmetry breaking *d*-wave would also produce the spectrum with a full gap feature. Because the V-shape spectra have been widely observed in our experiment, we thus believe that the *d*-wave gap should be a dominant one. And this is consistent with many theoretical calculations.

Intuitively, the pairing form in $Nd_{1-x}Sr_xNiO_2$ may serve as a bridge between the cuprate and the FeSC. Because the former has a single band feature and only the intra-orbital interaction as the driving force for pairing, leading to the *d*-wave gap; while the latter is a multiband system, one needs not only intra- but also the inter-orbital interaction for pairing, resulting in the orbital selective pairing and *s*-wave gaps with opposite signs on different Fermi pockets.[186-189] At the moment, we don't know whether the gaps on the hybridized-orbital-derived β and γ pockets are really fully gapped, and whether these gaps have opposite signs. The direct experimental evidence of the *d*-wave gap on the α pocket is also lacking. To resolve this issue, we need to do further phase-referenced quasiparticle interference experiments[190-193] on single-crystal samples when they are available, which has been conducted successfully in FeSC[186,194-196] and cuprates.[197] Clearly, more efforts are desired in order to pin down the assignment of the superconducting gaps on different Fermi pockets.

**OUTLOOK**

The discovery of infinite-layer nickelate superconductor marks the new area in the field of superconductivity. Here we make a brief review in this fast-developing field, focusing mainly on the electronic structures, magnetic excitations and superconducting pairing. Meanwhile, we also point out both similarities and distinctions between nickelates and cuprates.

Firstly, $R$NiO$_2$ is definitely of multiorbital feature, different from the single-band cuprate system. Due to the hybridizations between Ni $3d_{x^2-y^2}$ and R $5d$ orbitals, a small amount of holes are self-doped into the Ni $3d_{x^2-y^2}$ orbital, forming a dominant Fermi pockets originated from Ni $3d_{x^2-y^2}$ orbital and two extra small electron pockets derived mostly from $R$ $5d$ orbitals.

Secondly, the parent compound $R$NiO$_2$ locates in the regime of Mott insulator, one can naturally infer that the single occupied Ni $3d_{x^2-y^2}$ will easily form a local spin and AF long-range order. However, it shows no sign of any magnetic long-range order in the magnetic related measurements. Generally, the R $5d$ itinerant electrons can actively couple to Ni orbitals and screen the local spin of Ni, increasing the critical value of $U_{Ni}$ that is necessary to induce long-range magnetic ordering. Moreover, upon hole doping, the spin configuration of Ni $3d^8$ may be energetically arranged in two different ways, which depends on the limit whether the Hund's rule coupling is dominant over the crystal-field splitting (high-spin state, HS) or the opposite situation (low-spin state, LS). It remains highly debatable and worthy of more attention.

Thirdly, Pauli limited superconductivity with a very large value of Maki parameter α has been found in Nd$_{1-x}$Sr$_x$NiO$_2$ superconducting film, indicating unconventional pairing in the system. The tunneling spectra on Nd$_{1-x}$Sr$_x$NiO$_2$ thin films show predominantly two types of features, one has a V-shape, another a full gap. The measured V-shape spectra featuring a *d*-wave gap is the dominant pairing instability and is mainly originated from Ni-$3d_{x^2-y^2}$ orbital, which has been extensively demonstrated in many theoretical papers. Meanwhile, the full gap may have several possible explanations.

As for a future perspective, despite the discovery of superconductivity in all the three kinds of Nd, Pr, La based 112 systems, $T_c$ is still too low compared to that of cuprates. Strictly speaking, nickelates may not be categorized into the family of high temperature superconductors for now. We need to find out the reason and make effort to increase $T_c$. Secondly, bulk samples with the same structure and proper doping have no trace of superconductivity. It remains very elusive and needs enduring efforts toward resolving the problem. Concerning the phase diagram, we need to focus on whether there are emergent intertwined orders just like that in cuprates, especially for the relationship between AF spin fluctuations and superconductivity. The direct experimental evidence of the *d*-wave gap on α pocket is also lacking. With the development of sample synthesis and continuous accumulation of experimental results, we will find out the answers to these challenging issues.

To conclude, at the moment, it is just the beginning of research on superconductivity in the nickelate system. The phenomena observed so far show some similarities with cuprates, but more are distinctions. Thus the nickelate system provides a new platform for exploring unconventional superconductivity. The establishment and understanding of superconductivity in nickelates will shed new light on resolving the elusive pairing mechanism in high-temperature superconductors.

## ACKNOWLEDGENMETS

The authors are grateful to Harold Y. Hwang, A. Ariando, George A. Sawatzky, W. S. Lee, Fu-Chun Zhang, Kazuhiko Kuroki, Frank Lechermann, Xiangang Wan, Jinlong Yang, Weiqiang Yu, W. S. Lee, Ronny Thomale for useful discussions and some helps in wiring up this overview. This paper contains some of the published work from our group. We thank all the collaborators in those works. This work was supported by the National Key R&D Program of China (Grant nos. 2016YFA0300401 and 2016YFA0401704) and the National Natural Science Foundation of China (Grants:A0402/13001167, A0402/11774153, A0402/11534005, 1861161004, and A0402/11674164).

## AUTHOR CONTRIBUTIONS

Q. G. and H. H. W wrote the manuscript. All authors read and approved the final manuscript.

## DECLEARATION OF INTERESTS

The authors declare no competing interests.

## LEAD CONTACT WEBSITES

https://sc.nju.edu.cn.

# Figures

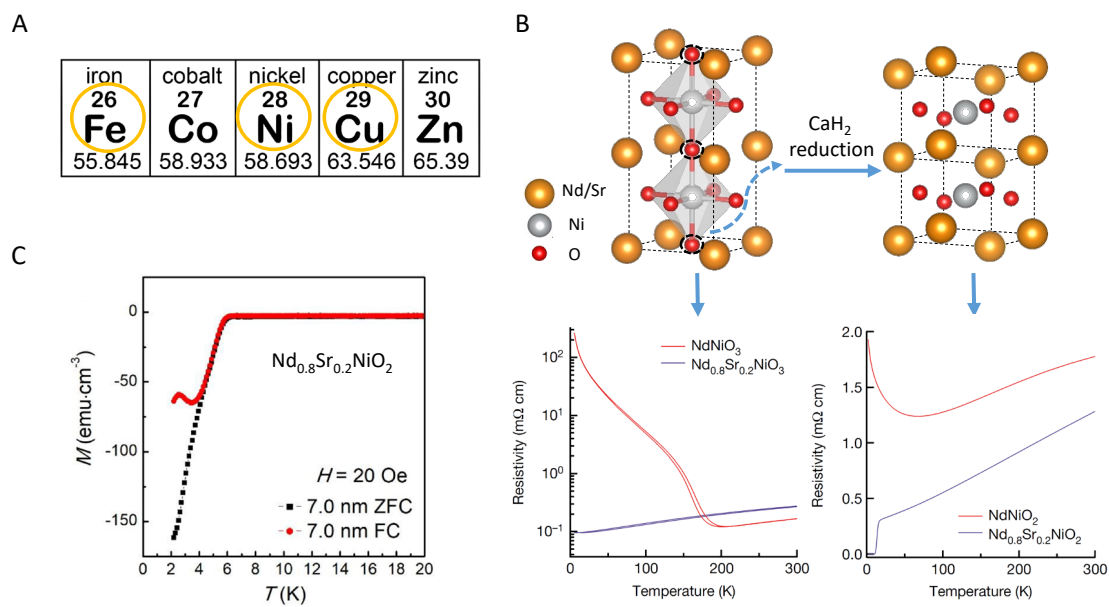

**Figure 1**. **Discovery of superconductivity in nickelate Nd$_{1-x}$Sr$_x$NiO$_2$ thin films**

(A) Some transition metals in the periodic table. Fe and Cu are two basic elements in two typical HTS; (B) Crystal structures of perovskite NdNiO$_3$ and infinite layer NdNiO$_2$. The shaded region shows that Ni is located at the center of NiO$_6$ octahedral. The corresponding temperature dependent resistivity is presented below. Adapted from Li et al.[46]; (C) ZFC and FC modes of DC magnetization on the thin film with typical thickness of 7 nm under the magnetic field of 20 Oe. Adapted from Zeng et al.[50]

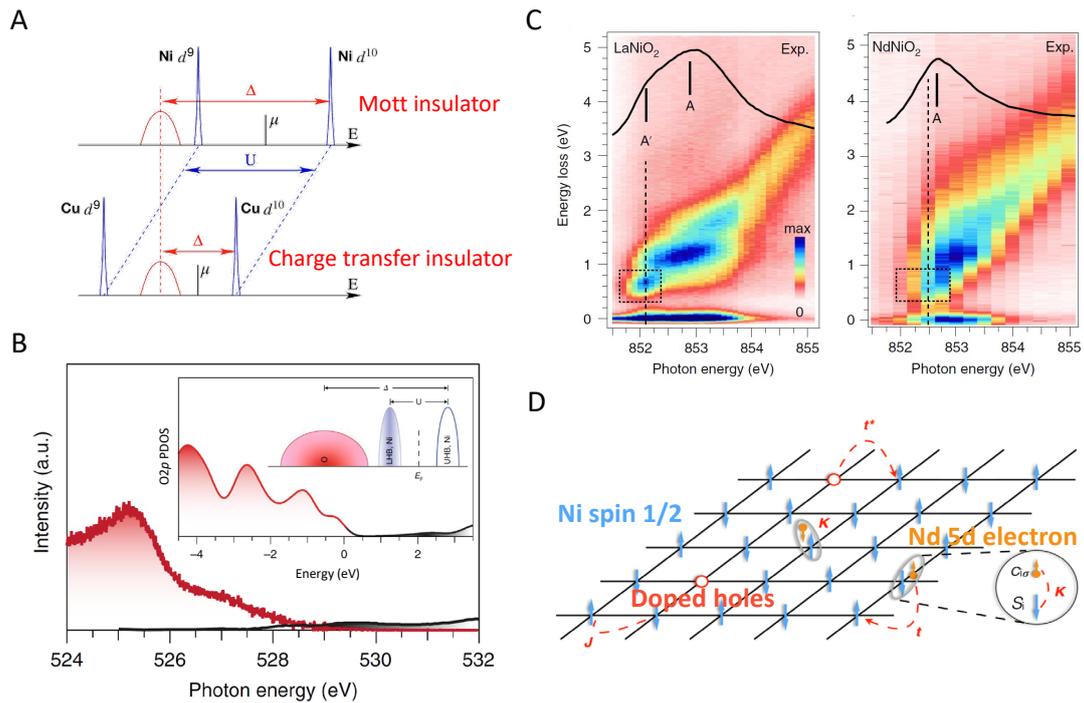

**Figure 2**. **Mott insulator, self-doping effect and possible Kondo coupling in nickelates $R$NiO$_2$**
(A) Sketch of a Mott insulator and a charge transfer insulator. The narrow (blue) bands are the Hubbard 3$d$ bands, while the broader (red) band is the O 2$p$ band before the $p$-$d$ hybridization. Adapted from Jiang et al.[134]; (B) XES and XAS in the pre-edge region of LaNiO$_2$, roughly reflecting the occupied (red shading) and unoccupied (black shading) oxygen PDOS, respectively. Inset: LDA + $U$ calculations for the PDOS with O 2$p$ orbital character and sketches of the relationship between $U$ and $\Delta$. UHB and LHB stand for upper and lower Hubbard band, respectively; (C) RIXS intensity map of LaNiO$_2$ and NdNiO$_2$. The corresponding XAS are superimposed as a solid black line in each map. The dashed boxes highlight the 0.6 eV features of LaNiO$_2$ and NdNiO$_2$ that are associated with the Ni–La and Ni–Nd hybridizations, respectively. Adapted from Hepting et al.[78]; (D) Illustration of the effective model on a two-dimensional square lattice of the NiO$_2$ plane of NdNiO$_2$. Blue arrow represents Ni spin, which interacts with its neighboring spin by AF coupling $J$. Orange arrow denotes a Nd 5$d$ electron, which couples to Ni spin by the Kondo coupling $K$, to form a Kondo singlet. Red circle represents Ni 3$d^8$ configuration, or a holon. The $t$ and $t^*$ are the hopping integrals of Kondo singlet and holon, respectively. Adapted from Zhang et al.[81]

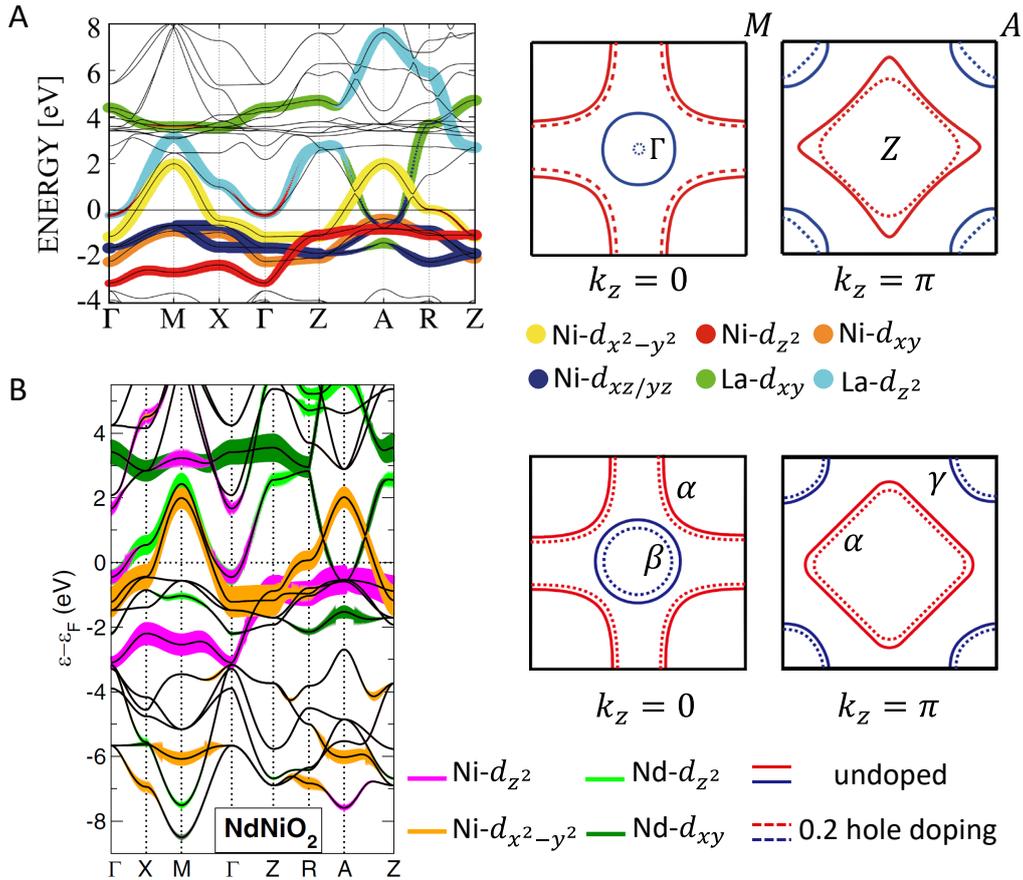

**Figure 3. Orbital resolve band structures of LaNiO$_2$ an NdNiO$_2$ in parent phase and 20% hole doping**

(A) Band structure of LaNiO$_2$ from the first-principles (solid lines). The band structure of the seven-orbital model is superposed, where the Wannier-orbital weight is represented by the thickness of lines with color-coded orbital characters. Top right-hand panels display cross sections of the Fermi surface at $k_z$ = 0 (left) and $k_z$ = π (right), where the red and blue lines depict Ni and La-originated Fermi surfaces, respectively. Adapted from Sakakibara et al.[97]; (B) DFT band structure of NdNiO$_2$. Undoped and 20% hole doped cases are plotted by solid lines and dashed lines, respectively. Adapted from Lechermann et al.[95]

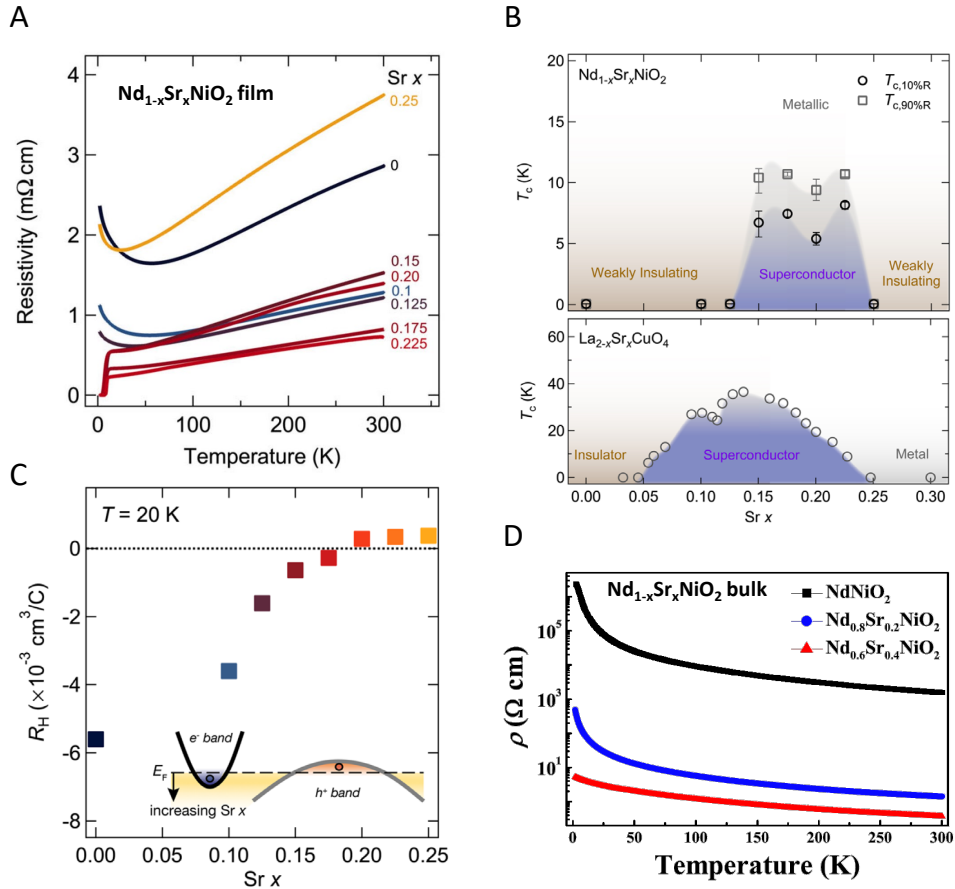

**Figure 4. Transport measurements and the resultant phase diagram**

(A) Temperature-dependent resistivity (2 K-300 K) measured for representative samples with different doping levels; (B) Top: phase diagram of $Nd_{1-x}Sr_xNiO_2$. Open circles (squares) represent $T_c$ of 10% $R$ ($T_c$ of 90% $R$), as defined to be the temperatures at which the resistivity is 10% (90%) of the resistivity value at 20 K. The symbols denote the average value across multiple samples of the same x. Bottom: phase diagram of $La_{2-x}Sr_xCuO_4$ with $T_c$ of 50% $R$ values; (C) $R_H$ as a function of x at 20 K, crossing zero between x = 0.175 and 0.2. The inset displays a two-band schematic (electron pocket, $e^-$ band; hole pocket, $h^+$ band). The arrow indicates the variation of the Fermi level $E_F$ with increasing x. Adapted from Li et al.[103]; (D) Temperature dependence of resistivity for $Nd_{1-x}Sr_xNiO_2$ (x = 0, 0.2, 0.4) bulk samples. Adapted from Li et al.[115]

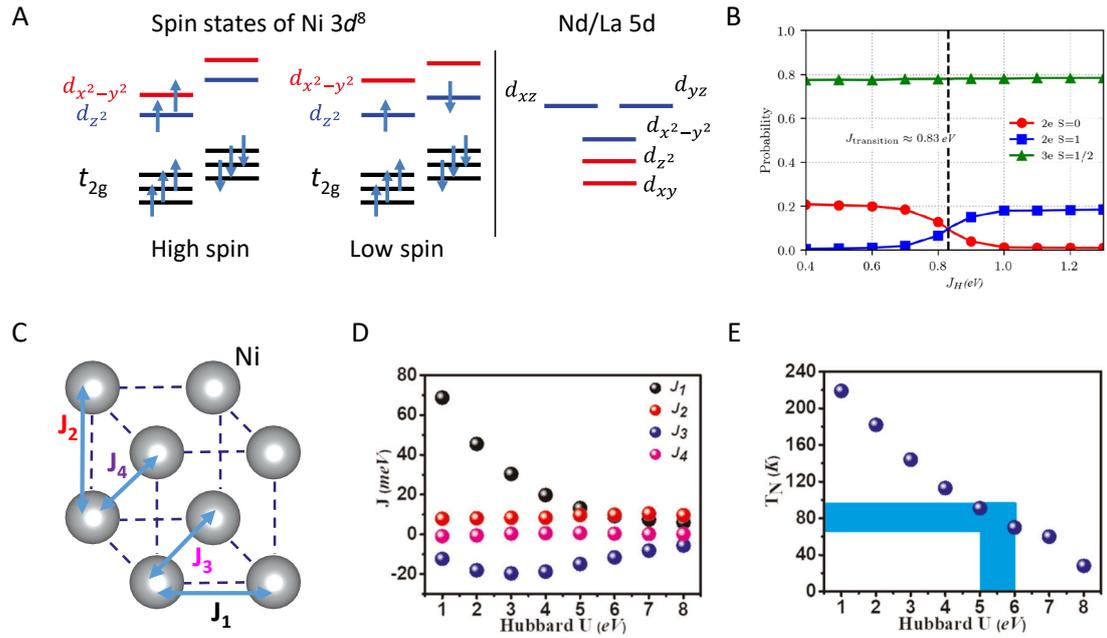

**Figure 5. Spin configurations and possible AF state**

(A) Left panel: spin configurations of Ni $3d^8$ in a square-planar environment. High spin state with $S = 1$ and low spin state with $S = 0$. Right panel: crystal field splitting on Nd; (B) calculated probabilities for the three-electron $S = 1/2$ and two-electron $S = 0$ and 1 states as a function of Hund's coupling $J_H$. Adapted from Wan et al.[88]; (C) The four exchange coupling parameters between Ni sites are indicated by the blue arrows; (D) the Hubbard $U$ dependent of exchange coupling parameters; (E) the estimated $T_N$. Adapted from Liu et al.[138]

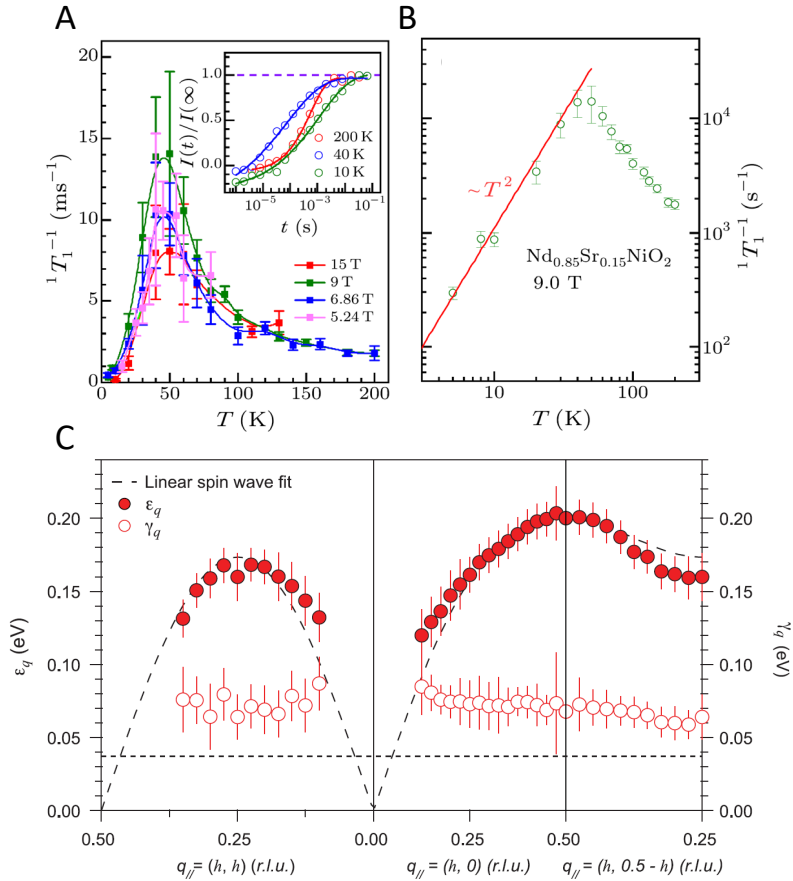

**Figure 6. Magnetic excitations**

(A) Temperature dependent $^1T_1^{-1}$ of $Nd_{0.85}Sr_{0.15}NiO_2$ under different fields. Inset: the spin recovery data at three typical temperatures. The solid lines are the exponential functions fitting to exract relaxation time $T_1$; (B) The temperature dependent $^1T_1^{-1}$ is plotted in a log-log scale, with a fitting of low temperature data to $^1T_1^{-1} \propto T^2$. Adapted from Cui et al.[144]; (C) Dispersion of magnetic excitations in $NdNiO_2$ and fit to the linear spin wave theory. Magnetic mode energy $\varepsilon$ is denoted by filled red circles and damping factor $\gamma$ is denoted by empty red circles, versus projected in-plane momentum transfers $q_{//}$ along high-symmetry directions. Adapted from Lu et al.[149]

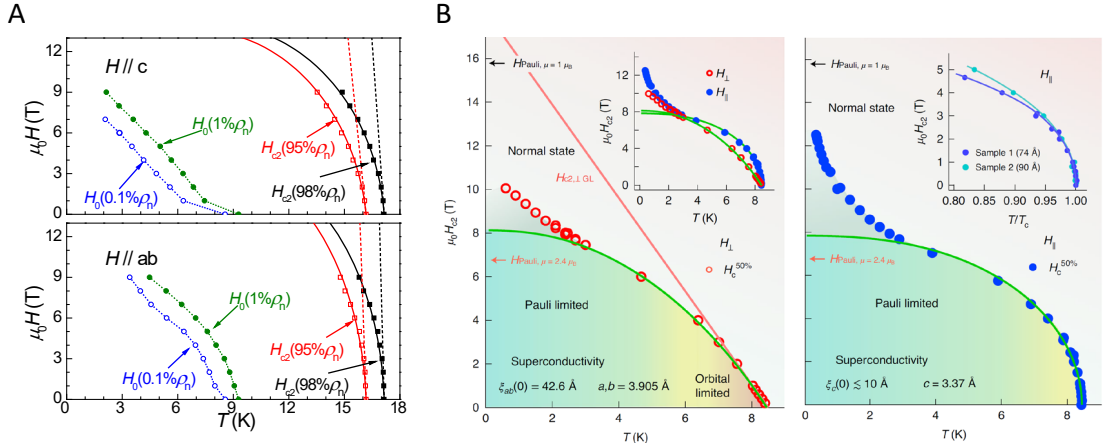

**Figure 7. Upper critical field and Pauli limited pairing**

(A) Superconducting phase diagram $H(T)$. Symbols used here denote the temperature-dependent $\mu_0 H_{c2}$ and $\mu_0 H_0$ obtained from $\rho$-$T$ curves, measured at different magnetic fields, and the solid (dashed) lines show the fitting results (theoretical curves) obtained from the WHH theory, from which the values of $\alpha$ can be extracted. $\mu_0 H_{c2}$ are determined using the criteria 95% $\rho_n(T)$, and 98% $\rho_n(T)$, $\mu_0 H_0$ are determined using the criteria 0.1% $\rho_n(T)$ and 1% $\rho_n(T)$. Adapted from Xiang et al.[153]; (B) Superconducting $H_{c2}$-$T$ phase diagrams for magnetic fields along the c axis and in the ab plane. The regions above $H_{c2}$ (open and filled circles) are shaded orange, representing the normal state, while the regions under $H_{c2}$ are shaded as a transition from red regions near $T_c$ to green regions near 0 K, representing a shift from superconductivity limited by orbital de-pairing to Pauli-limited superconductivity. The solid green lines are WHH fits of the $H_{c2}$ data above the low-temperature upturn. Adapted from Wang et al.[165]

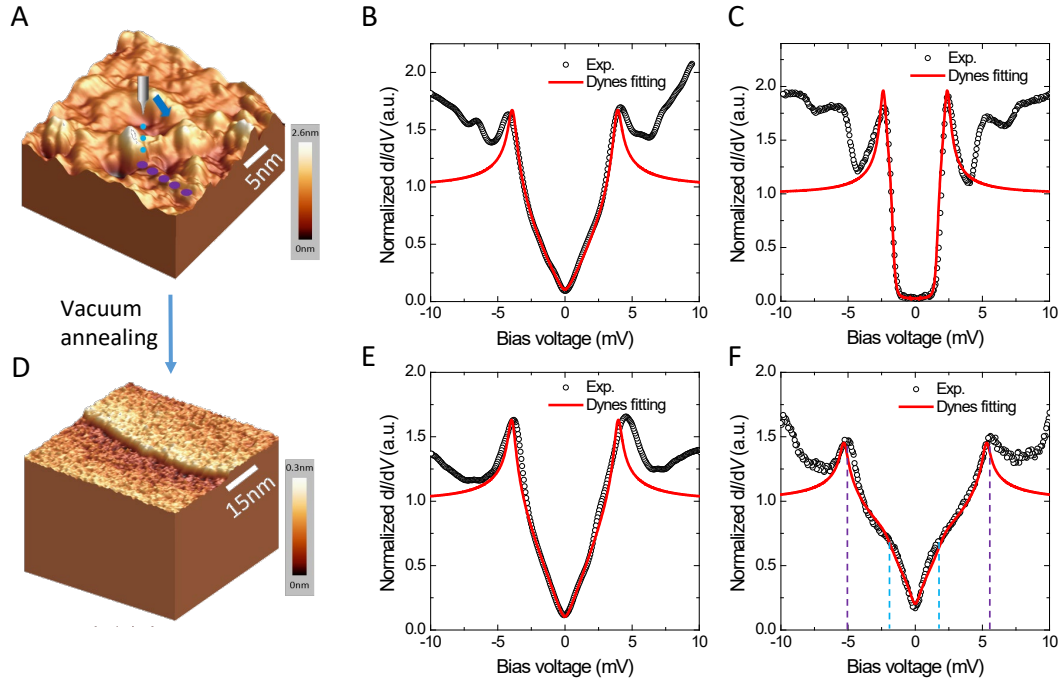

**Figure 8. STM measurement of topography and tunneling spectra**

(A) 3D illustration of the topographic image after top tactical reaction method, one can see the surface roughness about 1-2 nm; (B) A tunneling spectrum with a V-shape, which is measured at the rough surface. The Dynes model fitting yields a gap function $\Delta = 3.9\cos2\theta$ (meV); (C) A tunneling spectrum with full gap feature, which is measured at the rough surface. The Dynes model fitting is $\Delta = 2.35(0.85+0.15\cos4\theta)$ (meV); (D) 3D illustration of the topographic image after a long time vacuum annealing. The roughness becomes much smaller. We can see a clear step with the height about 0.17 nm, being consistent with one half of the unit cell height; (E) A typical V-shape spectrum measured on the smooth surface. A gap function $\Delta = 3.95(0.95\cos2\theta+0.05\cos6\theta)$ (meV) is used in the fitting; (F) A mixture of the two-gap features on the spectrum is measured on the smooth surface. Dynes fitting results, $\Delta_1 = 5.3(0.8\cos2\theta+0.2\cos6\theta)$ (meV), $\Delta_2 = 2$ meV, $p_1 = 85\%$ and $p_2 = 15\%$.

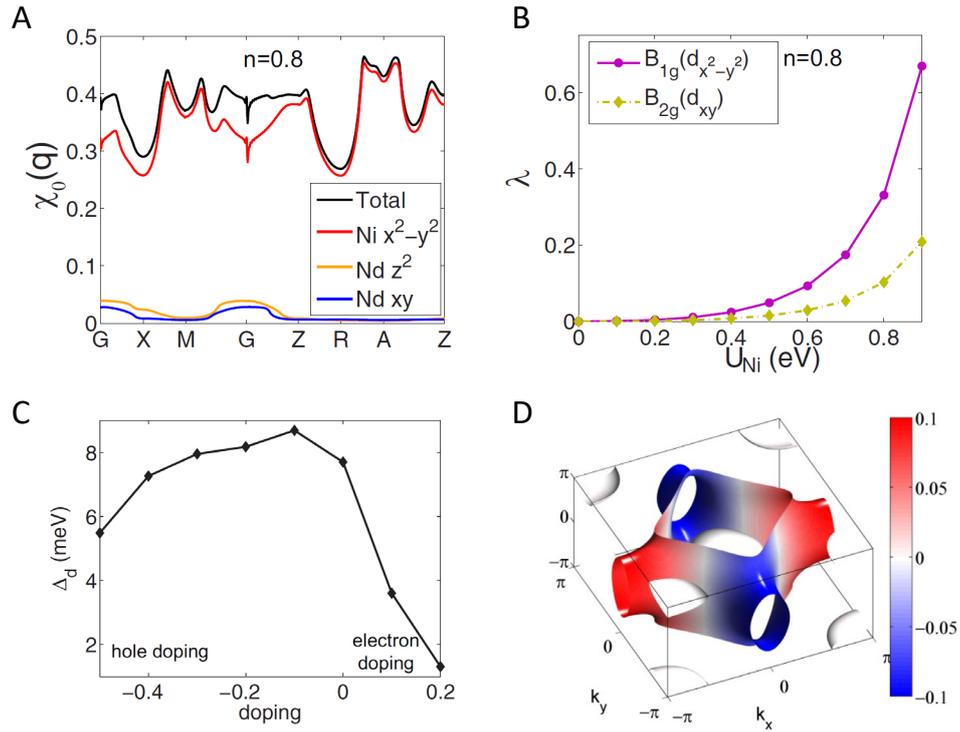

**Figure 9. Robust *d*-wave paring**

(A) Bare susceptibility for n = 0.8; (B) The pairing eigenvalues as a function of the interaction $U_{Ni}$ for n = 0.8; (C) The $d_{x^2-y^2}$-wave gap as a function doping with $J_1 = J_2 = 0.1$ eV. Positive (negative) values relate to electron (hole) doping; (D) 3D illustration of $d_{x^2-y^2}$-wave gap. Adapted from Wu et al.[90]

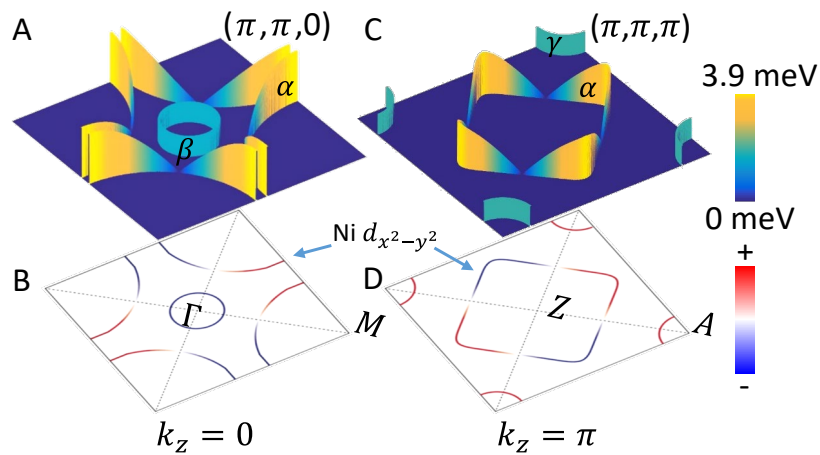

$\alpha$ : Ni-$3d_{x^2-y^2}$ **d wave**
$\beta, \gamma$ : hybridized orbitals, Ni-$3d_{3z^2-r^2}$/$d_{xz,yz}$ mixed with Nd-$5d_{3z^2-r^2}$/$d_{xy}$ **s wave**

**Figure 10. Cartoon picture for the Fermi surfaces and the gap structure**

(A),(B) The Fermi surfaces and the gap amplitude at cut $k_z = 0$. A *d*-wave gap is formed on the α Fermi pocket centered around Γ, and a full gap may appear on the β pocket. Actually the Fermi surface on the α pocket at this cut looks very similar to that of underdoped cuprates. The height of the colored walls in a represents gap magnitude on each Fermi surface; (C),(D) The Fermi surfaces and the gap amplitude at the cut $k_z = \pi$. A *d*-wave gap is formed on the α Fermi pocket centered around Z, now the Fermi surface becomes a closed square like mimicking that of overdoped cuprates, and a full gap may appear on the γ pocket around A. The height of the colored walls in (C) indicates the gap magnitude on each Fermi surface. The blue and red colors in (B) and (D) represent the gap signs. Adapted from Gu et al.[171]